# Unheard in the Digital Age:
# Rethinking AI Bias and Speech Diversity

**Onyedikachi Hope Amaechi-Okorie**
Technical Community Advocate, @JSON Schema, Nigeria
amaechihope20@gmail.com
Aula Fellowship for AI, Montreal, Canada
onyedikachi@theaulafellowship.org
ORCID: 0009-0004-0852-7971

**Branislav Radeljić** (Corresponding Author)
Aula Fellowship for AI, Montreal, Canada
branislav@theaulafellowship.org
ORCID: 0000-0002-0497-3470

**Abstract**
Speech remains one of the most visible yet overlooked vectors of inclusion and exclusion in contemporary society. While fluency is often equated with credibility and competence, individuals with atypical speech patterns are routinely marginalized. Given the current state of the debate, this article focuses on the structural biases that shape perceptions of atypical speech and are now being encoded into artificial intelligence. Automated speech recognition (ASR) systems and voice interfaces, trained predominantly on standardized speech, routinely fail to recognize or respond to diverse voices, compounding digital exclusion. As AI technologies increasingly mediate access to opportunity, the study calls for inclusive technological design, anti-bias training to minimize the impact of discriminatory algorithmic decisions, and enforceable policy reform that explicitly recognize speech diversity as a matter of equity, not merely accessibility. Drawing on interdisciplinary research, the article advocates for a cultural and institutional shift in how we value voice, urging co-created solutions that elevate the rights, representation, and realities of atypical speakers in the digital age. Ultimately, the article reframes speech inclusion as a matter of equity (not accommodation) and advocates for co-created AI systems that reflect the full spectrum of human voices.

## 1. Introduction

Speech is one of the most visible and audible ways humans participate in society. From classrooms to job interviews, and from podcasts to policy meetings, their voices often precede them, shaping how they are perceived (Ashby, 2011; Moss, 2020; Young, 2024). Yet, the systems that govern access to education, employment, leadership, and technology often exclude people whose voices do not align with dominant norms of fluency, clarity, or pace (Johnson, 2008; Klein & Hood, 2004; Miller 2015; Powers & Haller, 2017). In other words, speech remains a gatekeeper to credibility, participation, and power, whereby people with atypical speech styles, such as stuttering, cluttering, or neurodivergent patterns, are routinely misjudged as less competent, intelligent, or trustworthy, not for what they say but how they say it (Allard & Williams, 2008; Amick et al., 2017; Bailey et al., 2015). Internalized biases, often creating a feedback loop of stigma and underrepresentation, bear an immense emotional burden. Many individuals with atypical speech feel obliged to "mask" their natural patterns, avoiding speech situations, scripting conversations in advance, or hiding their difference altogether (Boyle & Cheyne, 2024; Hull et al., 2017; Plexico et al., 2009a). This emotional labor, coupled with constant pressure to "fix" their speech, leads to anxiety, exhaustion, and social withdrawal (Blood et al., 2011; Bosshardt, 2002; Boyle et al., 2023; Gerlach-Houck et al., 2023; Vanryckeghem et al., 2001).

While accessibility policies tend to address visible and auditory impairments, they overlook speech, further compounding its invisibility. This exclusion reflects broader systemic inequities shaped by ableism, colonial standards, and class-based notions of professionalism (Ayala-López, 2018). Technology has widened this gap. Automated speech recognition (ASR) systems and virtual assistants, often trained on fluent, standardized speech, fail to recognize atypical speech reliably, resulting in digital exclusion. When someone's voice is misunderstood or erased by the tools that increasingly mediate work, education, and civic participation, the message is clear: not all voices are valued equally. For those whose speech diverges from dominant norms, public speaking and spontaneous interaction become fraught with barriers imposed by institutional and social norms that dictate who gets validated in shared spaces (Clark et al., 2020; Rice & Kroll, 2006; Titchkosky, 2007). This suggests that addressing speech exclusion is not just about accessibility, but it is also about justice, since the recognition of atypical speech as a valid form of communication challenges the deep-rooted belief that fluency equals credibility or intelligence.

This article calls attention to the systemic barriers faced by individuals with atypical speech, including those with disfluent, neurodivergent, or accent-influenced patterns, and argues that their exclusion is not a personal issue but a political one. The lack of representation in leadership and public-facing roles only reinforces this marginalization. Rarely do we see people with non-fluent speech in positions of influence or portrayed authentically in media; instead, they are often framed as inspirational exceptions, celebrated not for their work but for "overcoming" how they speak. All this limits the cultural imagination and discourages others from fully participating in public life. Accordingly, this study calls for inclusive speech recognition systems trained on diverse speech data and co-designed with affected communities. In addition to facilitating policy change, it prompts a shift in how institutions define competence and participation. By making speech diversity visible in policy, technology, and public space, we create a more equitable world, one where everyone, regardless of how they speak, has the right to be heard, understood, and represented.

The article proceeds with an overview of the main debates surrounding atypical speech, both in traditional and technology-driven contexts. It then turns to the theoretical and methodological considerations that inform the present study and its contribution to the literature. Moving forward, the role of technology, including actors with the power to influence, if not directly shape, policy direction, is examined in greater detail. Their responsibilities and engagement with critical issues suggest that speech diversity is not a liability to be managed, but a capacity to be respected. In continuation, the paper explores three areas essential to fostering an inclusive digital voice culture: the use of voice technology itself, the elimination of speech-based discrimination, and the potential for improved policy and broader representation. These areas are expected to shed light on the overall maturity and preparedness to address the challenges of the digital age, as part of the general discourse on diversity, equity, inclusion, and accessibility (DEIA) policies. The article concludes by outlining the defining features of the politics of voice and offering insights for future research that may inform policymaking.

## 2. The state of the debate

The politics of voice determines who gets heard, who is taken seriously, and who is sidelined, not only in casual conversations, but also in policymaking, hiring decisions, and classroom participation (Ashby, 2011; Boyle et al., 2009; Butler, 2013; Isetti, 2014; Young, 2024). Kim and Lee (2013) demonstrate that teachers and peers often misjudge children who stutter as less capable. In addition, regardless of the context, such children are teased and bullied at school (Berchiatti et al., 2021). Both career advisors (Dew & Gabel, 2024) and hiring panels (Gabel et al., 2004; Hurst & Cooper, 1983) are less likely to recommend stuttering applicants for roles that involve communication skills. As explained by Kleinberg and Raghavan's (2018) model of implicit bias, interviewers unintentionally favor candidates who fit preconceived prototypes, and communication fluency is a key part of that prototype. All this translates into systemic bias in hiring and advancement (Gilman, 2012; Klein & Hood, 2004), as well as systematic under-employment, let alone removal from leadership opportunities (Abou-Dahech & Gabel, 2020; Boyle, 2016; Bricker-Katz et al., 2013; Gerlach et al., 2018; Yarzebinski, 2022).

The fact that atypical speech is tied to specific domains, not qualifications or expertise (Burrell-Kim, 2023), rarely appears in inclusion frameworks or accessibility policies, which tend to cast speech differences as personal challenges rather than valid forms of communication requiring structural support. Even more to the point, while inclusion and accessibility frequently address vision, hearing, and mobility, they often overlook speech diversity, including stuttering, accent-influenced speech, or neurodivergent verbal patterns entirely (Isaacs, 2020; Le, 2024; Lingras et al., 2023; Meredith & Packman, 2015; Morris, 2022). The absence of speech from formal guidelines reinforces a hierarchy of disability; in other words, if policy language does not name speech, then institutions rarely evaluate or resource support for it (Connery et al., 2020; Hamraie, 2017; Mack et al., 2021; Marshall et al., 2024; Titchkosky, 2011). Furthermore, in the absence of adequate frameworks and speech-aware environments, many individuals feel compelled to conceal their natural communication style (Gerlach-Houck et al., 2023; Wolfgruber et al., 2022). The emotional toll of losing touch with one's authentic voice is profound, as many internalize negative judgments and misattributions, blaming themselves rather than structural barriers for communication breakdowns (Azios et al., 2020; Boyle, 2015; Boyle et al., 2009; Johnson, 2008).

The inequities extend into technology, where voice assistants or ASR systems do not recognize atypical speech (Markl & McNulty, 2022; Mengesha et al., 2021; Mitra et al., 2021; Mujtaba et al., 2024b; Ngueajio & Washington, 2022). For example, many who stutter are cut off mid-command, misinterpreted by dictation tools, or forced to abandon voice interfaces altogether (Lea et al., 2023; Mujtaba & Mahapatra, 2025). Even with small improvements through fine-tuning, error rates remain unacceptably high (Wang et al., 2024). These failures are not isolated frustrations; they reflect how speech technology is designed. ASR systems are overwhelmingly trained on fluent, standardized speech, privileging certain accents and cadences while erasing others (Briggs & Thomas, 2015; Glasser et al., 2017; Shor et al., 2019; Tatman, 2017). Almost two decades ago, Rice and Kroll (2006) reported that over 70 percent of people who stutter believe their speech limits their employability, with more than half avoiding communication-heavy jobs due to stigma. This explains the more recent argument that inclusive voice design must explicitly account for atypical speech (Clark et al., 2020).

Regarding the question of true access, it begins when all communication styles are recognized as valid, and participation in public life is no longer contingent on fluency (Alper, 2017; Mujtaba et al., 2024b). Bencini (2023) emphasizes the importance of universal design approaches that anticipate variation from the start, rather than treating difference as a problem to retrofit. Dolata et al. (2022) stress that fairness in speech technology is a sociotechnical challenge, not a technical fix, and Treviranus (2018) reminds us that inclusion works best when systems expect variability from the start. Alongside this and to speed up the process, Sibanda and Mothapo (2024) recommend creating awareness by vigorous campaigns and addressing the inherent perceptions and stereotypes. Such a need is further justified by research observing that word error rates go up to 49 percent for users with severe stuttering, which points to the extent of exclusion characterizing the design of technical infrastructure (Lea et al., 2023). Apple's (2023) research highlights how poor performance directly correlates with user experiences of exclusion, while also demonstrating that targeted engineering can significantly improve accessibility. Therefore, the existing discrepancies are not simply technical oversights but reflections of broader social assumptions about what speech should sound like and who qualifies as a "typical" user (Ngueajio & Washington, 2022). Algorithms replicate the exclusion that already exists offline, turning voice technology into another gatekeeper (Mitra et al., 2021).

### 3. Theoretical and methodological considerations

Goodley (2014) argues that disability is too often located in the individual rather than in systems that refuse to accommodate difference. As Erevelles (2011) points out, dominant systems equate fluency with authority, speed with competence, and clarity with legitimacy. Put differently, disruptions such as pauses, repetitions, and disfluencies are routinely misread as incompetence or unreliability (Burgoon et al., 1990; Daniels & Gabel, 2004; Goodley, 2014). People with atypical speech are often seen through a distorted lens, mistaken for anxious, unreliable, or even deceptive, not because of what we say, but how we say it (Abasi, 2022; Gabel et al., 2004, Rana et al., 2021). Consequently, given that leaders emerge when they represent group prototypes (Hogg, 2001) and given that atypical speech is excluded from the "prototype," people who stutter or speak atypically are unlikely to be seen as leaders, even when they possess deep expertise or vision (Zeigler-Hill et al., 2020). While scripting conversations in advance, suppressing fidgeting or natural pauses, or even avoiding speaking situations entirely might smooth social interactions in the short term, such strategies come at a significant cost.

To begin with, one of the harmful outcomes involves chronic anxiety and exhaustion. The process of masking is not a neutral act; it demands constant monitoring of one's speech, anticipation of listener reactions, and quick adjustments to maintain fluency, all of which compound cognitive load (Bosshardt, 2006; Boyle, 2011; Büchel & Sommer, 2004; Cooper & Cooper, 1996). Over time, this sustained hypervigilance fosters chronic anxiety, as individuals brace themselves for possible breakdowns in communication or negative judgments (Craig et al., 2003; Craig et al., 2009). The emotional cost of staying "fluent enough" in educational and professional spaces has been linked to heightened stress responses and the onset of anxiety disorders among people who speak atypically (Blood et al., 2001; Boyle, 2013; Boyle et al., 2024; Iverach & Rapee, 2014). For students and professionals alike, weeks or months of continuous masking result in withdrawal, diminished authenticity, and a sense of alienation from both peers and work (Klompas & Ross, 2004). This pattern highlights how ableist norms around communication infiltrate daily life, making stuttering not just a speech difference but a chronic site of emotional labor (Hurst & Cooper, 1983). Therefore, as warned elsewhere, recognizing masking as emotional labor underscores the urgent need for systemic change in how institutions, workplaces, and communities approach speech diversity (Limura & Miyamoto, 2021).

Going forward, the issue of undermined self-worth is regular. The persistent pressure to perform fluency leads to an emotional labor that reframes atypical speech as a personal failure rather than a reflection of systemic ableism (Ashby, 2011; Bailey et al., 2015; Boyle, 2016; Daniels & Gabel, 2004; Gabel et al., 2004; Goodley, 2014). The weight of this internalized stigma, coupled with self-doubt and reduced willingness to participate, is reinforced by cultural and communicative norms that equate speech fluency with credibility and leadership potential (Burgoon et al., 1990; Zeigler-Hill et al., 2021). Educational and workplace environments often reproduce these hierarchies by framing speech diversity as a deficit to be corrected, rather than a legitimate form of communication (Butler, 2013; Dorsey & Guenther, 2000; Meredith & Packman, 2015; Rice & Kroll, 2006). This displacement of responsibility where the individual blames their own disfluency rather than the discriminatory context reinforces feelings of worthlessness and undermines identity (Connery et al, 2020; Daniels & Gabel, 2004).

In addition, the problem of loss of authentic voice should not be ignored either. The pressure to constantly monitor and "fix" speech turns conversation into performance rather than authentic expression (Plexico et al., 2009b). This shift strips away rhythm and spontaneity, reinforcing the idea that only fluent voices are valid in professional and social contexts (Boyle, 2013). Over time, fluency becomes prioritized over identity, as individuals are pressured to conform to dominant norms of communication (societal expectations) rather than being accepted for their natural voices (Bailey et al., 2015; Goodley, 2014). Accordingly, reclaiming authenticity, which goes hand in hand with trust, creativity, and leadership, requires shifting from deficit framings of speech toward recognizing stuttering and other disfluencies as valid forms of communication (Bencini, 2023; Clark et al., 2020).

Finally, reduced participation is also key, mostly because of the findings that many people who stutter or have other speech differences choose silence over exposure. This withdrawal is not rooted in lack of ability, but it is a survival strategy shaped by ableism and conversational norms that punish disfluent voices (Ayala-López, 2018; Connery et al., 2020; Goodley, 2014). Listeners often ascribe lower credibility and competence to non-normative speech, which can translate into professional penalties such as slower advancement and exclusion from leadership tracks (Allard & Williams, 2008; Amick et al., 2017; Lev-Ari & Keysar, 2010). To cope, many step back from projects that depend on verbal participation, limiting pathways that might otherwise elevate their careers (Klompas & Ross, 2004; Rice & Kroll, 2006). This disengagement fuels a cycle of invisibility, sustaining the belief that leadership and disfluency are incompatible and narrowing who is perceived as a credible leader (Dew & Gabel, 2024; Hogg, 2001; Wolfgruber et al., 2022).

With the above in mind, this article contributes to the literature by foregrounding speech diversity as a critical yet underexplored dimension of DEIA. While existing research has examined the social stigma and professional consequences of atypical speech, this study situates speech exclusion within broader systems of ableism, technological design, and policy neglect. It extends prior work by bridging disability studies, sociotechnical research, and policy analysis to show how speech-based discrimination is not merely a personal or interpersonal issue but a structural one, reinforced through education, employment practices, and algorithmic bias. Moreover, by emphasizing the emotional labor of masking and the systemic silencing of non-normative voices, the paper reframes fluency not as a neutral standard, but as a gatekeeping mechanism that restricts participation and leadership. It also advances the field by calling for inclusive voice technologies co-designed with affected communities, and by advocating for a shift in institutional definitions of competence. In doing so, the paper not only addresses a major gap in DEIA discourse but also proposes practical, justice-oriented interventions to ensure that all voices, regardless of fluency, are respected and represented.

## 4. Understanding the role of technology

Voice-activated technology is no longer a novelty, but the frontline of accessibility and participation (Larasati, 2025; Oumard et al., 2022). Still, misrecognitions are not just technical errors; they isolate the speaker in the moment, forcing them to repeat themselves or accept that their contribution will be mangled (Lev-Ari & Keysar, 2010; Mengesha et al., 2021). Over time, the pattern sends a message that such voices are not worth building for, which points to structural gatekeeping or a system designed and trained in ways that consistently sideline certain speech patterns (Markl & McNulty, 2022). Consequently, voice-based screening tools and automated hiring systems may reject or misinterpret atypical speech, further amplifying barriers (Hickman et al., 2024; Maindidze et al., 2025). Automated hiring tools have been reported to disadvantage individuals with disabilities through inaccessible designs based on standardized norms. When combined with speech biases, this results in a double exclusion for people with disfluent voices. Each failure compounds into reduced participation, lost opportunities, and the quiet removal of people from spaces where decisions are made (Buyl et al., 2022; Feng et al., 2024; Koenecke et al., 2020; Michel et al., 2025). In effect, ASR becomes a filter or hierarchical tool, granting seamless entry to some (or fluency and "neutral" speech), while leaving others stuck at the gate, speaking into a system that will not listen; in a way, it encodes the same biases that shape human discrimination, but with the speed, scaling, and seeming neutrality of a machine (Choi & Choi, 2025; DiChristofano et al., 2022; Mujtaba & Mahapatra, 2025; Slaughter et al., 2023).

A person who stutters while speaking with a regional or non-standard dialect navigates two simultaneous forms of bias, one toward disfluency, another toward accent or dialect variation, each a barrier on its own, but together creating a compounded exclusion that most voice technologies are not designed to address (Koenecke et al., 2020; Mujtaba et al., 2024a). These layers of bias do not operate in isolation; they interact in ways that intensify the harm (Markl & McNulty, 2022). A disfluent moment might already cause ASR to stumble, but when it occurs alongside dialectal pronunciation that the model was never trained on, the likelihood of misrecognition skyrockets (Gabler et al., 2023). The issue extends beyond accuracy to erasure, as transcripts may distort a speaker's words beyond recognition, replacing them with placeholders like "[inaudible]" or misattributing them entirely (Chowdhury et al., 2024), leaving a record that reflects the technology's limitations rather than the speaker's intended meaning. Looking more closely at the issue of disfluency, as Koenecke et al. (2020) put it, major ASR platforms transcribe African American Vernacular English with nearly double the error rate of Standard American English. The racial bias in voice technologies exposes the layered dimensions of speech accessibility (DiChristofano et al., 2022). Fluent speakers are not equally served if their dialect falls outside the trained norm and this represents a digital form of linguistic discrimination where only certain voices are heard and understood (Koenecke et al., 2020; Mengesha et al., 2021).

This compounded marginalization shapes behavior over time, as people, learning from repeated exclusion, may self-edit, avoid certain words, or adopt less natural speech patterns in an effort to "sound" more machine-readable (Mengesha et al., 2021). Moreover, following an experiment with 108 participants, Wenzel et al. (2023) concluded that when voice assistants produced high error rates, Black participants, unlike white participants, reported lower self-esteem, reduced positive affect, and less favorable evaluations of the technology. Users from African American communities in particular are inclined to report feelings of exclusion and they alter their own speech patterns by changing rhythm or word choice to fit what the technology could understand (Cohn et al., 2024). This adaptive effort exacts a cognitive and emotional toll, stripping spontaneity and authenticity from communication (Ritz et al., 2022). More specifically, users report they must alter their natural speech by slowing down, substituting words, or recalibrating pronunciation to work around the technology and this practice, often referred to as "voice masking" is not merely a technical workaround but a form of emotional labor (Lev-Ari & Keysar, 2010; Mujtaba et al., 2024a). Over time, this repeated effort can be exhausting, diverting cognitive energy away from the actual conversation and toward managing how one's voice will be received (Bosshardt, 2006; Jones et al., 2012). The toll of voice masking extends beyond fatigue. For many, it creates a subtle but persistent pressure to conform to a "standard" way of speaking (Giles et al., 1973), often one that aligns with the dialects and speech rhythms dominant in the model's training data (Serditova et al., 2025). For individuals whose voices carry regional dialects, non-native accents, or speech disfluencies, the choice becomes either to conform to the technology's limitations (Wu et al., 2020) or to accept reduced accuracy and, in some cases, exclusion from voice-enabled systems altogether (Mitra et al., 2021).

This dynamic also erodes trust in the technology itself, when users must repeatedly adjust themselves to accommodate a system, rather than the system adapting to them, they may question whether the technology truly serves their needs (Baughan et al., 2023). As Mengesha et al. (2021) observe, such interactions can leave users feeling unseen or undervalued, particularly when their speech patterns represent part of their identity. The act of changing one's voice to be "machine-readable" is not a neutral adaptation, but one that carries social and psychological weight, especially when tied to personal or cultural identity (Michel et al., 2025; Székely et al., 2025). In workplace or educational settings, the stakes are even higher. People who rely on voice interfaces for productivity or accessibility may find themselves expending disproportionate energy on making themselves understood (Lee et al., 2001). This energy is invisible in productivity metrics, yet it directly affects engagement, confidence, and overall performance (Esquivel et al., 2024). The cumulative effect is that those with non-standard speech bear an additional, unacknowledged cost in participating in technology-mediated spaces, a cost that could be mitigated by more inclusive and representative voice model design (Mengesha et al., 2021).

Bias is further reinforced through institutions of higher education, despite their positioning as meritocratic spaces where ability and achievement are meant to determine opportunity (Batruch et al., 2023; Dorsey & Guenther, 2000; Moriguchi et al., 2024; Naylor & Mifsud, 2019). Increasingly, these institutions rely on voice technologies to automate workflows, whether through lecture transcription software, oral examinations, voice-controlled classroom tools, or administrative systems that use ASR to process voice input. The promise of efficiency and accessibility quickly unravels when ASR systems fail to accurately transcribe a lecture given by a professor who stutters, or when a student's oral response is garbled in automated grading software (Kuhn et al., 2024). These failures are not minor technical glitches; they affect grades, evaluations, and access to accommodations (Ngueajio & Washington, 2022). In competitive environments, where academic recognition often translates into scholarships, research opportunities, and career advancement, even small errors compound into lasting inequities (Kizilcec & Lee, 2022). Technology, in this context, does not erase bias but encodes it into institutional processes (Yin et al., 2024). What makes this particularly insidious is that the bias is masked as impartial; when a student's speech is misrecognized, it is rarely the system that is blamed (Wenzel et al., 2023). Instead, responsibility is shifted back onto the speaker, reinforcing the idea that their voice is the problem (Goetsu & Sakai, 2020). This reflects broader patterns of institutional discrimination, where individuals are pressured to adapt to systems rather than systems adapting to human diversity, leaving students feeling compelled to mask their stutter, avoid speaking in class, or withdraw from oral components, further reducing their visibility and reinforcing stigma (Butler, 2013). In this way, voice technology becomes not a neutral tool of inclusion but an amplifier of institutional bias, determining whose knowledge and contributions are recorded, validated, and remembered (Choi & Choi, 2025).

## 5. Toward an inclusive digital voice culture
### 5.1. *Voice technology from exclusion to co-creation*

Automatic Speech Recognition (ASR) tools often fail people with accents, stutters, or other non-standard speech because they are trained on datasets that privilege fluent and normative voices (Koenecke et al., 2020; Tatman, 2017). Still, research demonstrates that ASR systems trained on speech from individuals with dysarthria, Parkinson's disease, or other conditions perform significantly better at recognizing non-normative speech (Shor et al., 2019; Wang et al., 2024). Thus, similar improvements could be expected in the context of datasets focusing on people who stutter, cluster, or speak with strong regional variations, which shifts performance from exclusion to meaningful participation.

However, data diversity is not just a technical requirement; it is an ethical obligation that ensures speech technologies reflect the full spectrum of human voices. With this in mind, inclusive datasets must be built in collaboration with speech-diverse communities, giving them agency over how their data is used and ensuring that participation results in tools that serve their needs (Briggs et al., 2015; Daniels & Gabel, 2004; Ngueajio & Washington; Treviranus, 2018). When people with lived experience are excluded from the design table, systems replicate the preferences of dominant voices while marginalizing those outside the norm (Craig et al., 2009; Goodley, 2014; Hamraie, 2017). Algorithmic fairness requires that accuracy metrics explicitly account for accented, disfluent, and atypical voices, otherwise technical progress continues to reproduce structural inequities (Mujtaba et al., 2024b). In addition, without transparent reporting standards, claims of "universal" usability obscure the fact that marginalized speakers continue to encounter disproportionate recognition errors (Koenecke et al., 2020).

Moving from exclusion to co-creation means designing datasets that celebrate difference rather than erase it, creating ASR systems that not only recognize diverse voices but affirm them as integral to our digital future. Indeed, fine-tuning general ASR systems with targeted, speech-specific datasets has shown significant reductions in error rates for stuttered and disfluent speech (Mujtaba et al., 2024b). Apple's (2023) Machine Learning Research demonstrated that specialized model adaptation meaningfully improved recognition accuracy for users with atypical speech patterns. Similarly, Lea et al. (2023) found that user-informed datasets improved usability and reduced frustration, directly linking technical optimization to lived experience. These findings confirm that personalization is not just a performance enhancement but a step toward equity in human–machine interaction (Ngueajio & Washington, 2022). By embedding fine-tuning as a standard design principle, the field moves from deficit-based framings of speech difference toward technological infrastructures that recognize and value multiplicity of voices (Bencini, 2023).

Calls for accountability in ASR development emphasize that inclusivity must be measured, not just promised (Dolata et al., 2022). When performance benchmarks are set exclusively around standardized, fluent speech, developers legitimize exclusionary systems that fail those with diverse communication needs (Tatman, 2017). Industry accountability also demands mechanisms for third-party auditing and community oversight in ASR evaluation (Ngueajio & Washington, 2022). The fact that user-led assessments reveal gaps that lab-based testing fails to capture, particularly when recognition errors disrupt professional participation and daily life, suggests that the absence of participatory accountability allows bias to persist unchecked, even as companies position themselves as inclusive innovators (Lea et al., 2023; Mengesha et al., 2021). Establishing accountability frameworks further requires alignment with disability justice principles that foreground systemic change over token adjustments, with standards being able to measure not only technical accuracy but also the extent to which technologies affirm communication rights and reduce stigma; by adopting accountability measures rooted in fairness and equity, developers move beyond efficiency narratives to address the sociocultural harms embedded in current systems (Bencini, 2023; Treviranus, 2018).

Lastly, accessibility must be reframed not as an accommodation but as a matter of justice that directly challenges ableism and exclusion. When voice technologies are designed only for standardized, fluent speech, they reinforce structural inequities and deny communication rights to those with speech differences (Bailey et al., 2015). This framing shifts the focus away from fixing individuals to dismantling barriers that delegitimize their voices in schools, workplaces, and digital spaces (Titchkosky, 2011). Therefore, recognizing accessibility as justice means acknowledging speech diversity as part of human variation (not a defect to be erased), but also demanding accountability from technology developers, who too often deploy speech recognition systems without considering exclusionary impacts (Ashby, 2011; Erevelles, 2011; Ngueajio & Washington, 2022; Sibanda & Mothapo, 2024). In other words, recognizing voice diversity as integral to universal design affirms communication rights and resists the silencing effects of linguistic ableism. This approach positions accessibility not as charity or compliance, but as the foundation of a digital future where everyone is equally heard.

### 5.2. Equitable hiring: Dismantling speech-based discrimination

Fluency bias often begins long before an interview, at the first stage of resume review where speech is assumed to signal credibility or competence, with studies showing that when communication is marked as "different," individuals are judged as less employable, which reinforces systemic patterns of exclusion and limited access (Allard & Williams, 2008; Amick et al., 2017; Gabel et al., 2004; Klein & Hood, 2004). Blind screening processes offer one way to disrupt this bias by anonymizing applications and limiting the influence of fluency-based assumptions at the earliest stage. Removing markers of speech or accent from hiring processes ensures candidates are evaluated on qualifications rather than stereotypes tied to verbal style (Boyle & Cheyne, 2024; Connery et al., 2020; Young, 2024). Hiring processes that move toward anonymity therefore shift the narrative, from managing "deficient" speakers to restructuring systems that have privileged fluency as a false measure of merit (Daniels & Gabel, 2004; Gilman, 2012).

Research shows that listeners frequently conflate fluency with intelligence, professionalism, or leadership potential, meaning that speech difference is unfairly taken as a lack of capability (Gabel et al., 2004; Gerlach et al., 2018), and that verbal performance is weighted more heavily than skill or experience (Burgoon et al., 1990; Lev-Ari & Keysar, 2010). Therefore, training interview panels to decouple speech difference from competence is essential to equitable hiring. Studies also highlight that structured exposure to individuals with atypical speech shifts workplace attitudes, reducing implicit bias and increasing fairness in evaluations (Iimura & Miyamoto, 2021; Sibanda & Mothapo, 2024). Therefore, equipping hiring teams with guidance on inclusive listening, patience in processing, and awareness of unconscious bias builds conditions where candidates are judged by their contributions rather than delivery. Accordingly, by recognizing speech diversity as part of human variation rather than deviation, interview processes can shift from exclusionary gatekeeping to genuine opportunity-making (Bricker-Katz et al., 2013; Wolfgruber et al., 2022).

For example, structured interview formats reduce subjective judgments and mitigate the snap impressions often triggered by speech difference (Kleinberg & Raghavan, 2018). Standardized questioning allows hiring decisions to rest on relevant competencies rather than on fluency or accent, aligning with the social model of disability that locates barriers in design rather than the individual (Goodley, 2014). By reducing opportunities for bias, structured formats interrupt the role entrapment frequently faced by people who stutter during evaluative encounters. At the same time, rigid structures without flexibility may inadvertently exclude candidates who need additional processing time or alternative communication formats (Ayala-López, 2018; Sibanda & Mothapo, 2024). Incorporating options such as extended pauses, written responses, or multimodal submissions can broaden access, ensuring that communication differences are not mistaken for a lack of competence (Bencini, 2023; Clark et al., 2020). In other words, embedding flexible accommodations into structured interviews moves organizations beyond compliance toward equity; when candidates are granted the conditions necessary to present themselves authentically, the hiring process shifts from gatekeeping to recognition of diverse communicative strengths (Connery et al., 2020; Dew & Gabel, 2024; Wolfgruber et al., 2022; Young, 2024).

In terms of legal aspects, people with diverse speech patterns have rights under disability frameworks, but those rights often go underutilized due to lack of awareness in HR and hiring systems; when HR teams are trained to anticipate and offer accommodations proactively, candidates are spared the additional burden of disclosure under pressure (Ayala-López, 2018; Goodley, 2014; Titchkosky, 2007). Such a support does not imply a favor but a legal obligation aligned with the social model of disability, which positions barriers in systems rather than in individuals (Goodley, 2014). Nonverbal options also disrupt the default assumption that verbal fluency equals credibility, an assumption rooted in ableist communication norms (Alper, 2017; Burgoon et al., 1990). These adjustments align with universal design principles, ensuring that hiring processes are not only compliant but structurally accessible to diverse candidates (Hamraie, 2017; Treviranus, 2018). Still, disclosure support is equally vital, as research shows candidates often conceal or downplay their speech difference for fear of discrimination, which compounds stress and reduces authenticity in interviews (Gabel et al., 2004; Gerlach-Houck et al., 2023). This goes hand in hand with the argument that training HR staff to respond with clarity, confidentiality, and actionable accommodations makes disclosure less risky and more empowering (Boyle et al., 2023; Dew & Gabel, 2024).

Lastly, public representation is not a superficial concern but a structural lever for cultural change, especially when individuals with atypical speech are visible in leadership positions (Azios et al., 2020; Daniels & Gabel, 2004; Johnson, 2008; Young, 2024; Zeigler-Hill et al., 2021). This is why campaigns such as "Stuttering Pride" are important, as they counter harmful framings by fostering collective identity and challenging inter-nalized stigma (Boyle et al., 2023). Positive media portrayals play a crucial role here, as seen in shifts when stories foreground resilience, expertise, and authenticity rather than pathology (Burrell-Kim, 2023; Kuster, 2011; Rasoli et al., 2025). Representation also works at the intersection of community and identity, providing role models that help individuals resist pressure to mask or overcompensate; seeing leaders with stutters visibly succeed can lessen internalized stigma and provide pathways for younger or early-career professionals to imagine themselves in similar roles (Blood et al., 2011; Berchiatti et al., 2021; Rana et al., 2021). This shift is critical in breaking cycles where communication difference is tied to self-doubt, making representation not symbolic but materially impactful (Craig et al., 2009). Without such models, the deficit narrative persists, further embedding ableist standards of speech into professional pipelines.

### *5.3. Policy and representation: Expanding the frame of inclusion*

Most accessibility policies omit speech, which renders millions of people effectively invisible and reflects a longstanding structural bias in disability discourse, where speech is often overlooked despite its central role in communication and participation (Ashby, 2011; Boyle et al., 2009; Boyle, 2016; Gabel et al., 2004; Titchkosky, 2007). In this case, explicit inclusion of speech diversity into accessibility frameworks is not only about systemic support, but also about culture and practice (Bencini, 2023;

Lingras et al., 2023; Marshall et al., 2024; Morris, 2022; Wolfgruber et al., 2022). Adding explicit language is also critical in a digital era where automatic speech recognition and voice interfaces mediate participation in work, healthcare, and daily life (Mujtaba et al., 2024b; Ngueajio & Washington, 2022). Without policy requirements, these technologies continue to exclude disfluent and non-normative speech, creating new layers of inequity; with this in mind, policies that explicitly cover speech diversity can mandate inclusive datasets, bias audits, and user-centered design, aligning technical systems with disability rights principles (Apple, 2023; Michel et al., 2025; Tatman, 2017).

While DEIA frameworks often emphasize race, gender, and physical disability, they rarely name speech differences explicitly. Therefore, including speech explicitly in DEIA statements establishes visibility across policy and practice, as well as organizational accountability in the design and use of technology (Lea et al., 2023; Treviranus, 2018). Institutional language has real consequences for how resources are allocated and what is prioritized within organizations (Hamraie, 2017; Wolfgruber et al., 2022;). Without naming speech, leaders lack the mandate to fund training, accommodation initiatives, or digital communication infrastructures (Apple, 2023; Alper, 2017; Morris, 2022; Titchkosky, 2011); conversely, naming speech diversity in DEIA statements ensures that datasets, auditing mechanisms, and co-design processes are developed with communication equity in mind (Ashby, 2011; Bencini, 2023).

Schools, businesses, and government agencies should integrate universal design principles into everyday communication practices, ensuring equitable participation across diverse speech styles (Bailey et al., 2015; Feng et al., 2024; Marshall et al., 2024; Mujtaba et al., 2024b). Studies in education show that flexible communication options improve learning outcomes broadly, highlighting how access initiatives often generate universal gains (Isaacs, 2020; Meredith & Packman, 2015;). Similarly, organizational research demonstrates that inclusive communication fosters more collaborative workplaces and reduces attrition linked to exclusion (Moss, 2020). When communication is designed to accommodate difference, equity shifts from a reactive adjustment to a proactive cultural standard (Hamraie, 2017). In the end, adopting inclusive communication practices aligns with broader movements for digital and workplace equity (Lea et al., 2023; Lingras et al., 2023; Michel et al., 2025). As ASR systems and voice technologies increasingly mediate access to services, excluding disfluent or accented speech compounds inequity, and, in this sense, inclusive communication is not an accommodation but a cornerstone of equitable public life.

The emotional toll of masking speech differences is profound, often creating a psychological strain that cannot be managed through grit or personal resilience alone (Bricker-Katz et al., 2013; Gerlach-Houck et al., 2023; Iverach & Rapee, 2014; Rana et al., 2021). This cycle perpetuates stigma by reinforcing the perception that fluent speech is the only acceptable standard (Tellis & St. Louis, 2015). The demand for masking, which is sustained by cultural narratives that link fluency with credibility, intelligence, and leadership potential (Zeigler-Hill et al., 2021), normalizes discrimination, pushing individuals into silence rather than encouraging inclusive listening environments (Amick et al., 2017). As a result, many choose to withdraw from public discourse, not because of an inability to contribute, but because participation requires an unsustainable emotional performance, which then reinforces societal myths that people with speech differences are less engaged or less capable (Berchiatti et al., 2021; Klompas & Ross, 2004). Based on such findings, design and technology must integrate inclusive practices that recognize and value diverse voices rather than treating them as errors (Mujtaba et al., 2024a).

Mainstream media often frame people who speak atypically through inspirational or deficit-driven narratives, perpetuating the expectation that speech must conform to dominant norms to be legitimate (Azios et al., 2020; Bailey et al., 2015; Burgoon et al., 1990; Johnson, 2008; Miller, 2015; Zeigler-Hill et al., 2021). As already explained elsewhere, correcting this imbalance requires a deliberate shift in how voices are represented across traditional and digital platforms (Burrell-Kim, 2023; Kuster, 2011; Powers & Haller, 2017; Rasoli Jokar et al., 2025). When diverse voices are integrated into political discourse and cultural production, audiences begin to decouple fluency from competence (Dolata et al., 2022; Sibanda & Mothapo, 2024). Representation must extend beyond "special interest," by involving CEOs, actors, and news anchors, to make speech diversity visible as part of human variation (Ashby, 2011;

Ayala-López, 2018; Bailey et al., 2015). This shift aligns with broader calls for inclusive design and equitable participation in digital and cultural life (Alper, 2017; Lea et al., 2023; Lingras et al., 2023; Wolfgruber et al., 2022).

Lastly, community-led organizations such as the British Stammering Association and the Indian Stammering Association play a critical role in reframing speech differences from deficits to valued forms of diversity. By challenging ableist narratives that measure competence by fluency, they create safe spaces where individuals can resist stigma, share coping strategies, and advocate for systemic change (Abasi, 2022; Boyle, 2016; Boyle et al., 2009; Connery et al., 2020; Sibanda & Mothapo, 2024). Funding for these groups is essential to ensure that advocacy is not limited to symbolic representation but backed with resources to influence policy and practice; without financial and institutional support, grassroots movements risk burnout and diminished reach, even as the communities they represent continue to face systemic discrimination (Marshall et al., 2024; also, Alper, 2017; Bencini, 2023; Klein & Hood, 2004; Moss, 2020). At the same time, these organizations are most effective when they remain community-driven rather than directed by external agendas (Batruch et al., 2023; Butler, 2013; Isaacs, 2020; Moriguchi et al., 2024). In fact, advocacy groups not only challenge stigma but also inform the design of inclusive technologies, which means that their involvement in co-design ensures that interventions are both technically effective and socially just (Briggs & Thomas, 2015; Dolata et al., 2022; Lea et al., 2023; Mujtaba et al., 2024b). Moreover, by connecting local struggles with global movements, these organizations position speech diversity as a civil rights issue with implications for equity across healthcare, education, and employment (Marshall et al., 2024; Morris, 2022).

### 6. Conclusion

The politics of voice is deeply embedded in the systems that govern access to power, inclusion, and opportunity. From hiring panels to speech recognition software, the world is built for those who conform to dominant speech norms (fluent, standardized, and uninterrupted). As pointed out, such norms are not neutral; they are shaped by historical, cultural, and institutional forces that systematically exclude people with atypical speech patterns, such as those who stutter, use non-dominant accents, or speak with neurodivergent rhythms.

As also pointed out, the issues are multifaceted. Individuals face emotional and professional burdens, while they feel forced to "mask" their natural patterns or compensate in silence. In workplaces, hiring processes often penalize non-fluency, with few formal mechanisms to address such bias. In technology, voice-driven AI systems, integral to everyday life frequently fail to recognize or respond to speech that does not match their training data. In media and leadership, people with atypical speech are rarely seen, limiting representation and reinforcing marginalization. And critically, in public policy, speech diversity remains absent or vaguely addressed, rendering it invisible in accessibility frameworks and rights-based approaches.

These are not isolated problems. They are systemic injustices with clear consequences such as digital exclusion, social marginalization, poor employability prospects, and so on. Yet, as this paper demonstrates, they are also addressable. The solutions exist; anti-bias training, inclusive speech datasets, participatory design processes, structured job interviews, enforceable legal provisions, and speech-inclusive policy language are not only possible, but necessary. At the core of these recommendations is a call to reframe speech diversity as a matter of equity and justice, instead of treating it as a matter of mere accommodation. In other words, by explicitly naming and addressing speech differences in accessibility laws, cultural narratives, and AI development, we begin to break the silence that surrounds it. Inclusion cannot be partial, but it must extend to how we listen, whose voices we amplify, and whose speech we design for.

Building on the foundations laid by this article, future research should expand the discourse on speech diversity across geographies, disciplines, and sectors. Scholars, but also activists and technologists, are invited to explore how speech norms intersect with race, class, gender, and neurodivergence, as well as how global and multilingual contexts further complicate notions of fluency and credibility in the digital age. More empirical studies are needed to document the lived experiences of individuals with atypical speech in different environments and to assess the effectiveness of existing interventions. Equally important is participatory research that centers the voices of those most affected, ensuring that policy, technology, and design processes are co-created rather than imposed. As AI systems continue to mediate access to opportunity, there is a critical window to shape the values embedded in these tools.

## References

Abasi, C.H. (2022). The science behind stuttering: Reducing stigma and public misconceptions. https://mavmatrix.uta.edu/honors_spring2022/56

Abou-Dahech, T., & Gabel, R. (2020). Vocational stereotyping of people who stutter: Human resource management students. *Perspectives of the ASHA Special Interest Groups*, *5*(5), 1139–1146. https://doi.org/10.1044/2020_PERSP-20-00003

Allard, E.R., & Williams, D.F. (2008). Listeners' perceptions of speech and language disorders. *Journal of Communication Disorders*, *41*(2), 108–123. https://doi.org/10.1016/j.jcomdis.2007.05.002

Alper, M. (2017). *Giving Voice: Mobile Communication, Disability, and Inequality*. Cambridge, MA: MIT Press.

Amick, L.J., Chang, S.E., Wade, J., & McAuley, J.D. (2017). Social and cognitive impressions of adults who do and do not stutter based on listeners' perceptions of read-speech samples. *Frontiers in Psychology*, *8*, 1148. https://doi.org/10.3389/fpsyg.2017.01148

Apple. (2023). Improved speech recognition for people who stutter. *Highlight*, May 18. https://machinelearning.apple.com/research/speech-recognition

Ashby, C.E. (2011). Whose "voice" is it anyway? Giving voice and qualitative research involving individuals that type to communicate. *Disability Studies Quarterly*, *31*(4). https://doi.org/10.18061/dsq.v31i4.1723

Ayala-López, S. (2018). A structural explanation of injustice in conversations: It's about norms. *Pacific Philosophical Quarterly*, *99*(4), 726–748. https://doi.org/10.1111/papq.12244

Azios, M., Irani, F., Rutland, B., Ratinaud, P., & Manchaiah, V. (2020). Representation of stuttering in the United Sates newspaper media. *Journal of Consumer Health on the Internet*, *24*(4), 329–345. https://doi.org/10.1080/15398285.2020.1810940

Bailey, K., Harris, S.J., & Simpson, S. (2015). Stammering and the social model of disability: Challenge and opportunity. *Procedia-Social and Behavioral Sciences*, *193*, 13–24. https://doi.org/10.1016/j.sbspro.2015.03.240

Batruch, A., Jetten, J., Van de Werfhorst, H., Darnon, C., & Butera, F. (2023). Belief in school meritocracy and the legitimization of social and income inequality. *Social Psychological and Personality Science*, *14*(5), 621–635. https://doi.org/10.1177/19485506221111017

Baughan, A., Wang, X., Liu, A., Mercurio, A., Chen, J., & Ma, X. (2023). A mixed-methods approach to understanding user trust after voice assistant failures. In *Proceedings of the 2023 CHI Conference on Human Factors in Computing Systems* (pp. 1–16). https://doi.org/10.1145/3544548.3581152

Bencini, G. (2023). Universal design and communication rights: Meeting the challenge of linguistic and communicative diversity. In *Design for Inclusion* (pp. 76–82). Amsterdam: IOS Press. https://doi.org/10.3233/shti230402

Berchiatti, M., Badenes-Ribera, L., Galiana, L., Ferrer, A., & Longobardi, C. (2021). Bullying in students who stutter: The role of the quality of the student–teacher relationship and student's social status in the peer group. *Journal of School Violence*, *20*(1), 17–30. https://doi.org/10.1080/15388220.2020.1812077

Blood, G. W., Blood, I.M., Tellis, G., & Gabel, R. (2001). Communication apprehension and self-perceived communication competence in adolescents who stutter. *Journal of Fluency Disorders*, *26*(3), 161–178. https://doi.org/10.1016/S0094-730X(01)00097-3

Blood, G.W., Blood, I.M., Tramontana, G.M., Sylvia, A.J., Boyle, M.P., & Motzko, G.R. (2011). Self-reported experience of bullying of students who stutter: Relations with life satisfaction, life orientation, and self-esteem. *Perceptual and Motor Skills*, *113*(2), 353–364. https://doi.org/10.2466/07.10.15.17.pms.113.5.353-364

Bosshardt, H.G. (2002). Effects of concurrent cognitive processing on the fluency of word repetition: Comparison between persons who do and do not stutter. *Journal of Fluency Disorders*, *27*(2), 93–114. https://doi.org/10.1016/s0094-730x(02)00113-4

Bosshardt, H.G. (2006). Cognitive processing load as a determinant of stuttering: Summary of a research programme. *Clinical Linguistics & Phonetics*, *20*(5), 371–385. https://doi.org/10.1080/02699200500074321

Boyle, M.P. (2011). Mindfulness training in stuttering therapy: A tutorial for speech-language pathologists. *Journal of Fluency Disorders*, *36*(2), 122–129. https://doi.org/10.1016/j.jfludis.2011.04.005

Boyle, M.P. (2013). Psychological characteristics and perceptions of stuttering of adults who stutter with and without support group experience. *Journal of fluency disorders*, *38*(4), 368–381. https://doi.org/10.1016/j.jfludis.2013.09.001

Boyle, M.P. (2015). Relationships between psychosocial factors and quality of life for adults who stutter. *American Journal of Speech-Language Pathology*, *24*(1), 1–12. https://doi.org/10.1044/2014_ajslp-14-0089

Boyle, M.P. (2016). Relations between causal attributions for stuttering and psychological well-being in adults who stutter. *International Journal of Speech-Language Pathology*, *18*(1), 1–10. https://doi.org/10.3109/17549507.2015.1060529

Boyle, M.P., Blood, G.W., & Blood, I.M. (2009). Effects of perceived causality on perceptions of persons who stutter. *Journal of Fluency Disorders*, *34*(3), 201–218. https://doi.org/10.1016/j.jfludis.2009.09.003

Boyle, M.P., & Cheyne, M.R. (2024). Major discrimination due to stuttering and its association with quality of life. *Journal of Fluency Disorders*, *80*, 106051. https://doi.org/10.1016/j.jfludis.2024.106051

Boyle, M.P., Cheyne, M.R., & Rosen, A.L. (2023). Self-stigma of stuttering: Implications for communicative participation and mental health. *Journal of Speech, Language, and Hearing Research*, *66*(9), 3328–3345. https://doi.org/10.1044/2023_jslhr-23-00098

Bricker-Katz, G., Lincoln, M., & Cumming, S. (2013). Stuttering and work life: An interpretative phenomenological analysis. *Journal of Fluency Disorders*, *38*(4), 342–355. https://doi.org/10.1016/j.jfludis.2013.08.001

Briggs, P., & Thomas, L. (2015). An inclusive, value sensitive design perspective on future identity technologies. *ACM Transactions on Computer-Human Interaction*, *22*(5), 1–28. https://doi.org/10.1145/2778972

Burgoon, J.K., Birk, T., & Pfau, M. (1990). Nonverbal behaviors, persuasion, and credibility. *Human communication research*, *17*(1), 140–169. https://doi.org/10.1111/j.1468-2958.1990.tb00229.x

Burrell-Kim, D. (2023). Stuttering Matt: Linguistic ableism and the mockery of speech impediments in video games. *Game Studies*, *23*(2). https://gamestudies.org/2302/articles/burrellkim

Butler, C. (2013). University?… hell no: Stammering through education. *International Journal of Educational Research*, *59*, 57–65. https://doi.org/10.1016/j.ijer.2013.03.002

Buyl, M., Cociancig, C., Frattone, C., & Roekens, N. (2022). Tackling algorithmic disability discrimination in the hiring process: An ethical, legal and technical analysis. In *Proceedings of the 2022 ACM Conference on Fairness, Accountability, and Transparency* (pp. 1071–1082). https://doi.org/10.1145/3531146.3533169

Büchel, C., & Sommer, M. (2004). What causes stuttering? *PLoS Biology*, *2*(2), e46. https://doi.org/10.1371/journal.pbio.0020046

Choi, A.S.G., & Choi, H. (2025). Fairness of automatic speech recognition: Looking through a philosophical lens. *arXiv*, *2508.07143*. https://doi.org/10.48550/arXiv.2508.07143

Chowdhury, P., Sarkar, N., Nath, S., & Sharma, U. (2024). Analyzing the effects of transcription errors on summary generation of Bengali spoken documents. *ACM Transactions on Asian and Low-Resource Language Information Processing*, *23*(9), 1–28. https://doi.org/10.1145/3678005

Clark, L., Cowan, B.R., Roper, A., Lindsay, S., & Sheers, O. (2020). Speech diversity and speech interfaces: Considering an inclusive future through stammering. In *Proceedings of the 2nd Conference on Conversational User Interfaces* (pp. 1–3). https://doi.org/10.1145/3405755.3406139

Cohn, M., Mengesha, Z., Lahav, M., & Heldreth, C. (2024). African American English speakers' pitch variation and rate adjustments for imagined technological and human addressees. *JASA Express Letters*, *4*(4). https://doi.org/10.1121/10.0025484

Connery, A., McCurtin, A., & Robinson, K. (2020). The lived experience of stuttering: a synthesis of qualitative studies with implications for rehabilitation. *Disability and Rehabilitation*, *42*(16), 2232–2242. https://doi.org/10.1080/09638288.2018.1555623

Cooper, E.B., & Cooper, C.S. (1996). Clinician attitudes towards stuttering: Two decades of change. *Journal of Fluency Disorders*, *21*(2), 119–135. https://doi.org/10.1016/0094-730X(96)00018-6

Craig, A., Blumgart, E., & Tran, Y. (2009). The impact of stuttering on the quality of life in adults who stutter. *Journal of Fluency Disorders*, *34*(2), 61–71. https://doi.org/10.1016/j.jfludis.2009.05.002

Craig, A., Hancock, K., Tran, Y., & Craig, M. (2003). Anxiety levels in people who stutter. *Journal of Speech, Language, and Hearing Research* *46*(5), 1197–1206. https://doi.org/10.1044/1092-4388(2003/093)

Daniels, D.E., & Gabel, R.M. (2004). The impact of stuttering on identity construction. *Topics in Language Disorders*, *24*(3), 200–215. https://doi.org/10.1097/00011363-200407000-00007

Dew, C.W., & Gabel, R.M. (2024). How perceived communication skills needed for careers influences vocational stereotyping of people who stutter. *Journal of Fluency Disorders*, *80*, 106039. https://doi.org/10.1016/j.jfludis.2024.106039

DiChristofano, A., Shuster, H., Chandra, S., & Patwari, N. (2022). Global Performance Disparities Between English-Language Accents in Automatic Speech Recognition. *arXiv*, *2208.01157*. https://doi.org/10.48550/arXiv.2208.01157

Dolata, M., Feuerriegel, S., & Schwabe, G. (2022). A sociotechnical view of algorithmic fairness. *Information Systems Journal*, *32*(4), 754–818. https://doi.org/10.1111/isj.12370

Dorsey, M., & Guenther, R.K. (2000). Attitudes of professors and students toward college students who stutter. *Journal of Fluency Disorders*, *25*(1), 77–83. https://doi.org/10.1016/S0094-730X(99)00026-1

Erevelles, N. (2011). *Disability and difference in global contexts: Enabling a transformative body politic*. New York: Palgrave Macmillan.

Esquivel, P., Gill, K., Goldberg, M., Sundaram, S. A., Morris, L., & Ding, D. (2024). Voice assistant utilization among the disability community for independent living: A rapid review of recent evidence. *Human Behavior and Emerging Technologies*, *1*, 6494944. https://doi.org/10.1155/2024/6494944

Feng, S., Halpern, B.M., Kudina, O., & Scharenborg, O. (2024). Towards inclusive automatic speech recognition. *Computer Speech & Language*, *84*, 101567. https://doi.org/10.1016/j.csl.2023.101567

Feng, S., Kudina, O., Halpern, B.M., & Scharenborg, O. (2021). Quantifying bias in automatic speech recognition. *arXiv*, *2103.15122*. https://doi.org/10.48550/arXiv.2103.15122

Gabel, R.M., Blood, G.W., Tellis, G.M., & Althouse, M.T. (2004). Measuring role entrapment of people who stutter. *Journal of Fluency Disorders*, *29*(1), 27–49. https://doi.org/10.1016/j.jfludis.2003.09.002

Gabler, P., Geiger, B.C., Schuppler, B., & Kern, R. (2023). Reconsidering read and spontaneous speech: Causal perspectives on the generation of training data for automatic speech recognition. *Information*, *14*(2), 137. https://doi.org/10.3390/info14020137

Gerlach, H., Totty, E., Subramanian, A., & Zebrowski, P. (2018). Stuttering and labor market outcomes in the United States. *Journal of Speech, Language, and Hearing Research*, *61*(7), 1649–1663. https://doi.org/10.1044/2018_JSLHR-S-17-0353

Gerlach-Houck, H., Kubart, K., & Cage, E. (2023). Concealing stuttering at school: "When you can't fix it… the only alternative is to hide it." *Language, Speech, and Hearing Services in Schools*, *54*(1), 96–113. https://doi.org/10.1044/2022_lshss-22-00029

Giles, H., Taylor, D.M., & Bourhis, R. (1973). Towards a theory of interpersonal accommodation through language: some Canadian data. *Language in society*, *2*(2), 177–192. https://doi.org/10.1017/S0047404500000701

Gilman, J. (2012). Disability or identity: Stuttering, employment discrimination, and the right to speak differently at work. *Brooklyn Law Review*, *77*(3), 1179. https://brooklynworks.brooklaw.edu/blr/vol77/iss3/7

Glasser, A., Kushalnagar, K., & Kushalnagar, R. (2017). Deaf, hard of hearing, and hearing perspectives on using automatic speech recognition in conversation. In *Proceedings of the 19th International ACM SIGACCESS Conference on Computers and Accessibility* (pp. 427–432). https://doi.org/10.1145/3132525.3134781

Goetsu, S., & Sakai, T. (2020). Different types of voice user interface failures may cause different degrees of frustration. *arXiv*, *2002.03582*. https://doi.org/10.48550/arXiv.2002.03582

Goodley, D. (2014). *Dis/ability Studies: Theorising Disablism and Ableism*. Oxon: Routledge.

Hamraie, A. (2017). *Building access: Universal design and the politics of disability*. Minneapolis, MN: University of Minnesota Press.

Hickman, L., Langer, M., Saef, R.M., & Tay, L. (2024). Automated speech recognition bias in personnel selection: The case of automatically scored job interviews. *Journal of Applied Psychology*, *110*(6), 846–858. https://doi.org/10.1037/apl0001247

Hogg, M.A. (2001). A social identity theory of leadership. *Personality and Social Psychology Review*, *5*(3), 184–200. https://doi.org/10.1207/S15327957PSPR0503_1

Hull, L., Petrides, K.V., Allison, C., Smith, P., Baron-Cohen, S., Lai, M.C., & Mandy, W. (2017). "Putting on my best normal:" Social camouflaging in adults with autism spectrum conditions. *Journal of Autism and Developmental Disorders*, *47*(8), 2519–2534. https://doi.org/10.1007/s10803-017-3166-5

Hurst, M.I., & Cooper, E.B. (1983). Employer attitudes toward stuttering. *Journal of Fluency Disorders*, *8*(1), 1–12. https://doi.org/10.1016/0094-730X(83)90017-7

Iimura, D., & Miyamoto, S. (2021). Public attitudes toward people who stutter in the workplace: A questionnaire survey of Japanese employees. *Journal of Communication Disorders*, *89*, 106072. https://doi.org/10.1016/j.jcomdis.2020.106072

Isaacs, D. (2020). "I don't have time for this:" Stuttering and the politics of university time. *Scandinavian Journal of Disability Research*, *22*(1), 58–67. https://doi.org/10.16993/sjdr.601

Isetti, D.D. (2014). *Listener impressions of spasmodic dysphonia: Symptom severity and disclosure of diagnosis during a job interview* (Doctoral dissertation, University of Washington). https://digital.lib.washington.edu/server/api/core/bitstreams/bdc2ff4b-3766-477a-8a66-ca0022c4fe15/content

Iverach, L., & Rapee, R.M. (2014). Social anxiety disorder and stuttering: Current status and future directions. *Journal of Fluency Disorders*, *40*, 69–82. https://doi.org/10.1016/j.jfludis.2013.08.003

Johnson, J.K. (2008). The visualization of the twisted tongue: Portrayals of stuttering in film, television, and comic books. *Journal of Popular Culture*, *41*(2), 245. https://doi.org/10.1111/j.1540-5931.2008.00501.x

Jones, R.M., Fox, R.A., & Jacewicz, E. (2012). The effects of concurrent cognitive load on phonological processing in adults who stutter. *Journal of Speech, Language, and Hearing Research*, *55*(6), 1862–1875. https://doi.org/10.1044/1092-4388(2012/12-0014)

Kim, S.-M., & Lee, E.-J. (2013). A qualitative study on teachers' perceptions and response processes towards preschool children who stutter. *Communication Sciences & Disorders*, 18(2): 203–222. https://doi.org/10.12963/csd.13020

Kizilcec, R.F., & Lee, H. (2022). Algorithmic fairness in education. In *The Ethics of Artificial Intelligence in Education* (pp. 174–202). Oxon: Routledge.

Klein, J.F., & Hood, S.B. (2004). The impact of stuttering on employment opportunities and job performance. *Journal of Fluency Disorders*, *29*(4), 255–273. https://doi.org/10.1016/j.jfludis.2004.08.001

Kleinberg, J., & Raghavan, M. (2018). Selection problems in the presence of implicit bias. *arXiv, 1801.03533*. https://doi.org/10.48550/arXiv.1801.03533

Klompas, M., & Ross, E. (2004). Life experiences of people who stutter, and the perceived impact of stuttering on quality of life: Personal accounts of South African individuals. *Journal of Fluency Disorders*, *29*(4), 275–305. https://doi.org/10.1016/j.jfludis.2004.10.001

Koenecke, A., Nam, A., Lake, E., Nudell, J., Quartey, M., Mengesha, Z., ... & Goel, S. (2020). Racial disparities in automated speech recognition. *Proceedings of the National Academy of Sciences*, *117*(14), 7684–7689. https://doi.org/10.1073/pnas.1915768117

Kuhn, K., Kersken, V., Reuter, B., Egger, N., & Zimmermann, G. (2024). Measuring the accuracy of automatic speech recognition solutions. *ACM Transactions on Accessible Computing*, *16*(4), 1–23. https://doi.org/10.1145/3636513

Kuster, J.M. (2011). At long last, a positive portrayal of stuttering. *The ASHA Leader*, *16*(2), 13–25. https://doi.org/10.1044/leader.FTR2.16022011.13

Larasati, R. (2025). Inclusivity of AI speech in healthcare: A decade look back. *arXiv, 2505.10596*. https://doi.org/10.48550/arXiv.2505.10596

Le, L. (2024). "I am human, just like you:" What intersectional, neurodivergent lived experiences bring to accessibility research. In *Proceedings of the 26th International ACM SIGACCESS Conference on Computers and Accessibility* (pp. 1–20). https://doi.org/10.1145/3663548.3675651

Lea, C., Huang, Z., Narain, J., Tooley, L., Yee, D., Tran, D. T., ... & Findlater, L. (2023). From user perceptions to technical improvement: Enabling people who stutter to better use speech recognition. In *Proceedings of the 2023 CHI Conference on Human Factors in Computing Systems* (pp. 1–16). https://doi.org/10.1145/3544548.3581224

Lee, J.D., Caven, B., Haake, S., & Brown, T.L. (2001). Speech-based interaction with in-vehicle computers: The effect of speech-based e-mail on drivers' attention to the roadway. *Human Factors*, *43*(4), 631–640. https://doi.org/10.1518/001872001775870340

Lev-Ari, S., & Keysar, B. (2010). Why don't we believe non-native speakers? The influence of accent on credibility. *Journal of Experimental Social Psychology*, *46*(6), 1093–1096. https://doi.org/10.1016/j.jesp.2010.05.025

Lingras, K.A., Alexander, M.E., & Vrieze, D.M. (2023). Diversity, equity, and inclusion efforts at a departmental level: Building a committee as a vehicle for advancing progress. *Journal of Clinical Psychology in Medical Settings*, *30*(2), 356–379. https://doi.org/10.1007/s10880-021-09809-w

Mack, K., McDonnell, E., Jain, D., Lu Wang, L., E. Froehlich, J., & Findlater, L. (2021). What do we mean by "accessibility research?" A literature survey of accessibility papers in CHI and ASSETS from 1994 to 2019. In *Proceedings of the 2021 CHI Conference on Human Factors in Computing Systems* (pp. 1–18). https://doi.org/10.1145/3411764.3445412

Maindidze, H.T., Randall, J.G., Martin‐Raugh, M.P., & Smith, K.M. (2025). A meta-analysis of accent bias in employee interviews: The effects of gender and accent stereotypes, interview modality, and other moderating features. *International Journal of Selection and Assessment*, *33*(1), e12519. https://doi.org/10.1111/ijsa.12519

Markl, N., & McNulty, S.J. (2022). Language technology practitioners as language managers: Arbitrating data bias and predictive bias in ASR. *arXiv, 2202.12603*. https://doi.org/10.48550/arXiv.2202.12603

Marshall, J., Wylie, K., McLeod, S., McAllister, L., Barrett, H., Owusu, N.A., ... & Atherton, M. (2024). Communication disability in low and middle-income countries: A call to action. *BMJ Global Health*, *9*(7), e015289. https://doi.org/10.1136/bmjgh-2024-015289

Mengesha, Z., Heldreth, C., Lahav, M., Sublewski, J., & Tuennerman, E. (2021). "I don't think these devices are very culturally sensitive:" The impact of errors on African Americans in automated speech recognition. *Frontiers in Artificial Intelligence*, *4*. https://doi.org/10.3389/frai.2021.725911

Meredith, G., & Packman, A. (2015). The experiences of university students who stutter: A quantitative and qualitative study. *Procedia: Social and Behavioral Sciences*, *193*, 318–319. https://doi.org/10.1016/j.sbspro.2015.03.293

Michel, S., Kaur, S., Gillespie, S.E., Gleason, J., Wilson, C., & Ghosh, A. (2025). "It's not a representation of me:" Examining accent bias and digital exclusion in synthetic AI voice services. In *Proceedings of the 2025 ACM Conference on Fairness, Accountability, and Transparency* (pp. 228–245). https://doi.org/10.1145/3715275.3732018

Miller, T. (2015). Stuttering in the Movies: Effects on Adolescents' Perceptions of People who Stutter. https://mabel.wwu.edu/do/13f2eb0c-7877-4b6d-a724-83c184edeb2e

Mitra, V., Huang, Z., Lea, C., Tooley, L., Wu, S., Botten, D., ... & Bigham, J. (2021). Analysis and tuning of a voice assistant system for dysfluent speech. *arXiv, 2106.11759*. https://doi.org/10.48550/arXiv.2106.11759

Moriguchi, C., Narita, Y., & Tanaka, M. (2024). Meritocracy and its discontents: Long-run effects of repeated school admission reforms. *arXiv,*

Morris, M.A. (2022). Striving toward equity in health care for people with communication disabilities. *Journal of Speech, Language, and Hearing Research*, *65*(10), 3623–3632. https://doi.org/10.1044/2022_jslhr-22-00057

Moss, H. (2020). Screened out onscreen: Disability discrimination, hiring bias, and artificial intelligence. *Denver Law Review*, *98*(4), 775–805. https://dx.doi.org/10.2139/ssrn.3906300

Mujtaba, D., & Mahapatra, N.R. (2025). Fine-tuning ASR for stuttered speech: Personalized vs. generalized approaches. *arXiv, 2506.00853*. https://doi.org/10.21437/Interspeech.2025-2373

Mujtaba, D., Mahapatra, N.R., Arney, M., Yaruss, J.S., Gerlach-Houck, H., Herring, C., & Bin, J. (2024a). Lost in transcription: Identifying and quantifying the accuracy biases of automatic speech recognition systems against disfluent speech. *arXiv, 2405.06150*. https://doi.org/10.48550/arXiv.2405.06150

Mujtaba, D., Mahapatra, N.R., Arney, M., Yaruss, J.S., Herring, C., & Bin, J. (2024b). Inclusive ASR for disfluent speech: Cascaded large-scale self-supervised learning with targeted fine-tuning and data augmentation. *arXiv, 2406.10177*. https://doi.org/10.48550/arXiv.2406.10177

Naylor, R., & Mifsud, N. (2019). Structural inequality in higher education: Creating institutional cultures that enable all students. Australian Centre for Student Equity and Success. https://www.acses.edu.au/app/uploads/2019/09/NaylorMifsud-FINAL.pdf

Ngueajio, M.K., & Washington, G. (2022). Hey ASR system! Why aren't you more inclusive? Automatic speech recognition systems' bias and proposed bias mitigation techniques. A literature review. In *International Conference on Human-Computer Interaction* (pp. 421–440). Berlin: Springer. https://doi.org/10.1007/978-3-031-21707-4_30

Oumard, C., Kreimeier, J., & Götzelmann, T. (2022). Pardon? An overview of the current state and requirements of voice user interfaces for blind and visually impaired users. In *International Conference on Computers Helping People with Special Needs* (pp. 388–398). Cham: Springer. https://doi.org/10.1007/978-3-031-08648-9_45

Plexico, L.W., Manning, W.H., & Levitt, H. (2009a). Coping responses by adults who stutter: Part I. Protecting the self and others. *Journal of Fluency Disorders*, *34*(2), 87–107. https://doi.org/10.1016/j.jfludis.2009.06.001

Plexico, L.W., Manning, W.H., & Levitt, H. (2009b). Coping responses by adults who stutter: Part II. Approaching the problem and achieving agency. *Journal of Fluency Disorders*, *34*(2), 108–126. https://doi.org/10.1016/j.jfludis.2009.06.003

Powers, E.M., & Haller, B. (2017). Journalism and mass communication textbook representations of verbal media skills: Implications for students with speech disabilities. *Journal of Media Literacy Education*, *9*(2), 58–75. https://doi.org/10.23860/JMLE-2019-09-02-05

Rana, H.J., Kausar, R., & Khan, N. (2021). Social anxiety and quality of life: Mediating role of stigma perception in individuals who stutter. *NUST Journal of Social Sciences and Humanities*, *7*(2), 232–246. https://doi.org/10.51732/njssh.v7i2.92

Rasoli Jokar, A.H., Karimi, H., & Yaruss, J.S. (2025). Stuttering representation on X: A detailed analysis of content, sentiment, and influences. *American Journal of Speech-Language Pathology*, *34*(4), 2156–2169. https://doi.org/10.1044/2025_ajslp-24-00375

Rice, M., & Kroll, R. (2006). The impact of stuttering at work: Challenges and discrimination. In *International Stuttering Awareness Day Online Conference*. https://ahn.mnsu.edu/services-and-centers/center-for-communication-sciences-and-disorders/services/stuttering/professional-education/convention-materials/archive-of-online-conferences/isad2006/

Ritz, H., Wild, C.J., & Johnsrude, I.S. (2022). Parametric cognitive load reveals hidden costs in the neural processing of perfectly intelligible degraded speech. *Journal of Neuroscience*, *42*(23), 4619–4628. https://doi.org/10.1523/jneurosci.1777-21.2022

Serditova, D., Tang, K., & Steffens, J. (2025). Automatic speech recognition biases in Newcastle English: An error analysis. *arXiv, 2506.16558*. https://doi.org/10.48550/arXiv.2506.16558

Shor, J., Emanuel, D., Lang, O., Tuval, O., Brenner, M., Cattiau, J., ... & Matias, Y. (2019). Personalizing ASR for dysarthric and accented speech with limited data. *arXiv, 1907.13511*. https://doi.org/10.48550/arXiv.1907.13511

Sibanda, R., & Mothapo, T. C. (2024). Communicative practices and perceptions towards stuttering people in South Africa. *South African Journal of Communication Disorders*, *71*(1), 1008. https://doi.org/10.4102/sajcd.v71i1.1008

Slaughter, I., Greenberg, C., Schwartz, R., & Caliskan, A. (2023). Pre-trained speech processing models contain human-like biases that propagate to speech emotion recognition. *arXiv, 2310.18877*. https://doi.org/10.48550/arXiv.2310.18877

Székely, É., & Miniota, J. (2025). Will AI shape the way we speak? The emerging sociolinguistic influence of synthetic voices. *arXiv, 2504.10650*. https://doi.org/10.48550/arXiv.2504.10650

Tatman, R. (2017). Gender and dialect bias in YouTube's automatic captions. In *Proceedings of the First ACL Workshop on Ethics in Natural Language Processing* (pp. 53–59). https://doi.org/10.18653/v1/W17-1606

Tellis, G., & St Louis, K.O. (2015). *Stuttering Meets Stereotype, Stigma, and Discrimination: An Overview of Attitude Research*. Morgantown, WV: West Virginia University Press.

Titchkosky, T. (2007). *Reading and Writing Disability Differently: The Textured Life of Embodiment*. Toronto, ON: University of Toronto Press.

Titchkosky, T. (2011). *The Question of Access: Disability, Space, Meaning*. Toronto, ON: University of Toronto Press.

Treviranus, J. (2018). *The three dimensions of inclusive design: A design framework for a digitally transformed and complexly connected society* (Doctoral dissertation, University College Dublin). https://openresearch.ocadu.ca/id/eprint/2745/1/TreviranusThesisVolume1&2_v5_July%204_2018.pdf

Vanryckeghem, M., Hylebos, C., Brutten, G.J., & Peleman, M. (2001). The relationship between communication attitude and emotion of children who stutter. *Journal of Fluency Disorders*, *26*(1), 1–15. https://doi.org/10.1016/S0094-730X(00)00090-5

Wang, H., Jin, Z., Geng, M., Hu, S., Li, G., Wang, T., ... & Liu, X. (2024). Enhancing pre-trained ASR system fine-tuning for dysarthric speech recognition using adversarial data augmentation. In *2024 IEEE International Conference on Acoustics, Speech and Signal Processing (ICASSP)* (pp. 12311–12315). https://doi.org/10.1109/ICASSP48485.2024.10447702

Wenzel, K., Devireddy, N., Davison, C., & Kaufman, G. (2023). Can voice assistants be micro-aggressors? Cross-race psychological responses to failures of automatic speech recognition. In *Proceedings of the 2023 CHI Conference on Human Factors in Computing Systems* (pp. 1–14). https://doi.org/10.1145/3544548.3581357

Wolfgruber, D., Stürmer, L., & Einwiller, S. (2022). Talking inclusion into being: Communication as a facilitator and obstructor of an inclusive work environment. *Personnel Review*, *51*(7), 1841–1860. https://doi.org/10.1108/PR-01-2021-0013

Wu, Y., Rough, D., Bleakley, A., Edwards, J., Cooney, O., Doyle, P.R., ... & Cowan, B.R. (2020). See what I'm saying? Comparing intelligent personal assistant use for native and non-native language speakers. In *22nd International Conference on Human–Computer Interaction with Mobile Devices and Services* (pp. 1–9). https://doi.org/10.1145/3379503.3403563

Yarzebinski, C.S. (2022). *An Exploration of the Management of Stuttering During Job Interviews*. Lafayette, LA: University of Louisiana at Lafayette.

Yin, Z., Chinta, S.V., Wang, Z., Gonzalez, M., & Zhang, W. (2024). FairAIED: Navigating fairness, bias, and ethics in educational AI applications. *arXiv, 2407.18745*. https://doi.org/10.48550/arXiv.2407.18745

Young, M.M. (2024). *Stuttering in the Workplace: Employers' Attitudes, Employability Ratings, and Hiring Decisions* (Doctoral dissertation, University of Texas at Austin). https://repositories.lib.utexas.edu/server/api/core/bitstreams/76e69dd1-d139-460a-9bff-1fff44498ed2/content

Zeigler-Hill, V., Besser, A., & Besser, Y. (2021). The negative consequences of stuttering for perceptions of leadership ability. *Journal of Individual Differences*, *42*(3), 116–123. https://psycnet.apa.org/doi/10.1027/1614-0001/a000336